\documentclass{aastex}
\usepackage{emulateapj5}

\def\kms{km~s$^{-1}$}

\def\ga{\mathrel{\hbox{\rlap{\hbox{\lower4pt\hbox{$\sim$}}}\hbox{$>$}}}}
\def\la{\mathrel{\hbox{\rlap{\hbox{\lower4pt\hbox{$\sim$}}}\hbox{$<$}}}}

\shorttitle{SST observations of SNR 1E 0102.2-7219}

\shortauthors{S.\ Stanimirovi\'{c} et al.}

\begin{document}

\title{{\it Spitzer Space Telescope} detection of the young supernova
  remnant 1E 0102.2-7219}

\author{Sne\v{z}ana Stanimirovi\'{c}$^{1}$, Alberto D. Bolatto$^{1}$, Karin
  Sandstrom$^{1}$, Adam K. Leroy$^{1}$, Joshua D. Simon$^{1,2}$,
  B. M. Gaensler$^{3}$, Ronak Y. Shah$^{4}$, James M. Jackson$^{4}$}
\setcounter{footnote}{1}
\footnotetext{Radio Astronomy Lab, UC Berkeley, 601 Campbell Hall,
Berkeley, CA 94720; 
(sstanimi, bolatto, karin, aleroy, jsimon)@astro.berkeley.edu}
\setcounter{footnote}{2}
\footnotetext{Current address: Department of 
Astronomy, California Institute of Technology, 1200 E. California Blvd, MS 
105-24, Pasadena, CA  91125}
\setcounter{footnote}{3}
\footnotetext{Harvard-Smithsonian Center for Astrophysics, 60 Garden Street, 
Cambridge, MA 02138; bgaensler@cfa.harvard.edu}
\setcounter{footnote}{4}
\footnotetext{Institute for Astrophysical Research, 
Boston University, 725 Commonwealth Avenue, Boston, MA 02215; 
(ronak, jackson)@bu.edu}

\begin{abstract}
We present infrared observations of the young, oxygen-rich 
supernova remnant 1E 0102.2-7219 (E0102) in the Small Magellanic Cloud,
obtained with the {\it Spitzer Space Telescope}.
The remnant is detected at 24 $\mu$m but not at 8 or 70 $\mu$m 
and has a filled morphology with two prominent filaments. 
We find evidence for the existence of up to 
$8\times10^{-4}$ M$_{\odot}$ of hot dust ($T_{d}\sim120$ K) 
associated with the remnant.
Most of the hot dust is located in the central region of E0102 which
appears significantly enhanced in infrared and radio continuum
emission relative to the X--ray emission.
Even if {\it all} of the hot dust was formed in the explosion of E0102,
the estimated mass of dust is at least 100 times lower that what 
is predicted by some recent theoretical models. 
\end{abstract} 


\section{Introduction}

Young supernova remnants (SNRs) are exciting laboratories for studying 
energetic phenomena occurring in the interstellar
medium (ISM) such as cosmic ray acceleration, nucleosynthesis, dust
formation and destruction, and collisionless shock physics
\citep{McKee01,Vink04}. One small subclass of young SNRs, with less
than a dozen members, is the so called oxygen--rich SNRs.
These remnants have very high--velocity debris ($V>1000$ \kms) and
high abundances of oxygen, neon, carbon, and magnesium. 
They are believed to be remnants of the most massive stars
(typical mass of $>15$ M$_{\odot}$). The oxygen-rich SNRs arise mainly
from Type Ib/c supernovae, and their typical representative is Cas
A. Another very young oxygen-rich SNR, found in the Small Magellanic
Cloud (SMC), is 1E 0102.2-7219 (hereafter E0102), 
commonly considered as dynamically very similar to Cas A.

E0102 is a well studied object at optical, radio, and X--ray
wavelengths. 
Its kinematic age is approximately 1000 years, 
making it one of the youngest SNRs known.
E0102 is located $\sim15$~pc in projected distance NE 
of the edge of the massive star--forming region LHA 115-N 76C (N~76,
Henize 1956). 
The remnant is a prominent radio source at 6 cm, with
a shell-like morphology and a diameter of $40\arcsec$ \citep{Amy93}.  
X--ray images of E0102 show a faint, filled
circular structure that marks the location of hot gas associated with
the forward moving blast wave. 
The bright X--ray ring --- mainly due to strong emission 
lines of O, Ne and Mg --- marks the
interaction of the reverse shock with the stellar ejecta and shows
significant substructure: a bright knot in
the southwest with a radial extension or ``spoke'', a bright arc in
the southeast, and a bright linear feature or ``shelf'' to the north
\citep{Gaetz00,Flanagan04}. 
All observations point to a very massive progenitor star ($>20$ M$_{\odot}$).

We report here the detection of E0102 at 24 $\mu$m 
in the {\it Spitzer Space Telescope (SST)} Survey of the 
SMC\footnote{http://celestial.berkeley.edu/spitzer} (S$^{3}$MC).
We summarize observations and data reduction
strategy in \S 2. The infrared (IR) morphology of E0102 is presented in \S 3, 
and compared with observations at other wavelengths in \S 4. We discuss the
origin of the IR emission associated with the remnant in \S 5, and
summarize our results in \S 6.

\section{Observations and Data Processing}
\label{s:obs}

Images presented here were obtained as a part of the S$^{3}$MC program that
images the SMC in all Infrared Array Camera (IRAC) and Multiband
Imaging Photometer (MIPS) bands (Bolatto et al., in preparation). 
The {\it SST} mosaics were constructed using the MOPEX data reduction
package \citep{Makovoz05}.
The angular resolution is 2\arcsec, 6\arcsec and 18\arcsec, at 8, 24 
and 70 $\mu$m, respectively.
All images were flat-fielded to remove the residual background.
For the 24 $\mu$m image, a fourth order polynomial 
function was modeled for each image row around the SNR and subtracted to 
properly separate the IR emission associated with the remnant 
from the surrounding diffuse emission associated with N76.
The final images have a noise level of 
0.07, 0.08, and 0.5 MJy sr$^{-1}$ at 8, 24, and 70 $\mu$m,
respectively.

\section{1E 0102.2-7219 at 24 and 70 $\mu$m}

\begin{figure*}
\epsscale{1.5}
\plotone{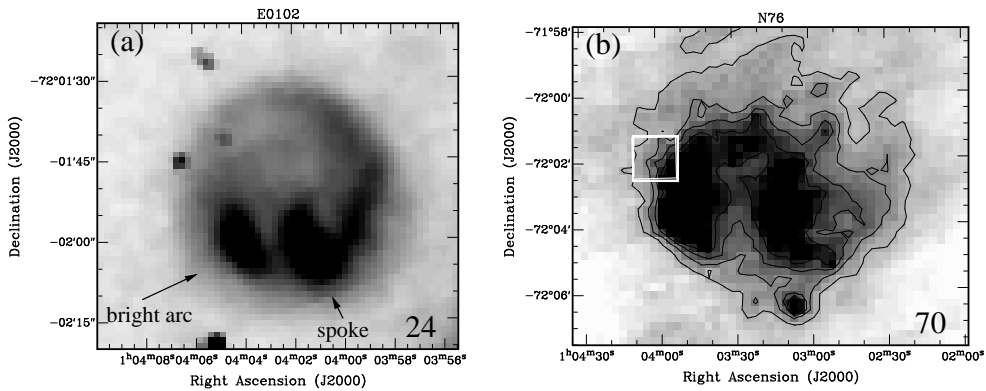}
\caption{\label{f:24m_70m} 
(a) MIPS 24 $\mu$m image of E0102, obtained by the {\it SST}. 
The gray scale intensity range is 0 to 3.2 MJy sr$^{-1}$
with a linear transfer function. 
Several dark pixels around the SNR are affected by cosmic rays.
(b) MIPS 70 $\mu$m image of N 76. The gray scale intensity range 
is 0 to 25 MJy sr$^{-1}$ with a linear transfer function; 
contours range from 3.7 to 4.9 MJy sr$^{-1}$, 
with a step of 0.25 MJy sr$^{-1}$.
The white box represents the area shown at 24 $\mu$m (a).
A small bump seen in two contours inside the white box 
(exactly where the SNR is located) may indicate a
possible small contribution from the remnant.}
\end{figure*}

E0102 is prominent in the 24 $\mu$m {\it SST} image, while it is not
clearly detected in any of the other {\it SST} bands at the depth of our
observations.  The 24 $\mu$m MIPS image of E0102 is shown in
Figure~\ref{f:24m_70m}a.  The remnant displays an
almost filled morphology, with some limb brightening along the west
side, and two bright elongated knots of emission to the south. 
The two bright knots, with a mean intensity of $\sim3$ 
MJy~sr$^{-1}$, have major axes parallel to each other.  
The bigger and slightly brighter knot to the west
corresponds to the ``spoke'' feature seen in X--rays, while the smaller
knot to the east is spatially coincident with the bright arc seen in the X-ray
ring \citep{Gaetz00,Flanagan04}.

Conversely, at 70 $\mu$m and with the resolution provided by the {\it SST}
(18\arcsec), it is not easy to extricate the remnant from the extended
IR emission from the nearby \ion{H}{2} region N~76. 
Figure~\ref{f:24m_70m}b shows the MIPS image of N 76 at 70 $\mu$m,
where the white box represents the area around E0102 
that is shown in Figure~\ref{f:24m_70m}a.
E0102 is not obviously detected at 70
$\mu$m, although there is a small bump present in the low-level IR
contours exactly at the remnant's position.
The measured flux density of the whole remnant at 24 $\mu$m is $80\pm4$
mJy, while 1-$\sigma$ upper limits at 8 and 70 $\mu$m are 3 
and 20 mJy, respectively. The resultant flux density ratios are 
$S_{70}/S_{24}\la0.25$, and $S_8/S_{24}\la0.04$.

\section{Comparison with X-ray and radio continuum data}

\begin{figure*}
\epsscale{1.5}
\plotone{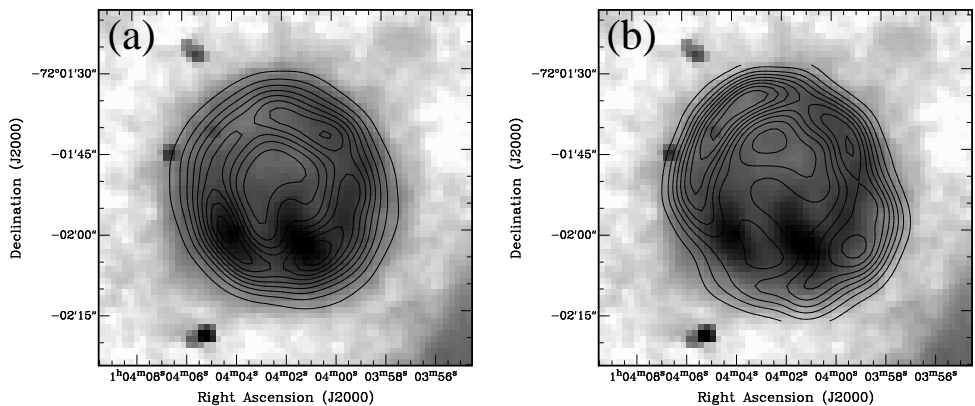}
\caption{\label{f:24m_Xray} (a) The 24 $\mu$m image of
E0102 overlaid with the {\it Chandra} X--ray (0.3--10 keV) contours from
\cite{Gaetz00}.  The X--ray image was smoothed to match the resolution
of the IR image. The gray scale intensity range is $-0.4$ to 4.5
MJy sr$^{-1}$, with a logarithmic transfer function;
the X-ray contours range over 10--90\% of the peak value,
with a step of 10\%. (b) The 24 $\mu$m image of
E0102 overlaid with the radio continuum contours at 6 cm from
\cite{Amy93}. The radio image was smoothed to match the resolution
of the IR image; radio contours range from 17.7 to 159.3 mJy beam$^{-1}$,
with a step of 17.7 mJy beam$^{-1}$.}
\end{figure*}

Observations of young SNRs at different wavelengths 
trace different regions/stages of the SNR/ISM interaction.
The X--ray emission arises from both forward and reverse shocks.
The hotter outer ring identifies the forward shock, while
the bright, inner ring originates from the reverse-shocked stellar ejecta,
and is seen in various emission lines \citep{Hayashi94,Flanagan04}. 
The 6 cm radio continuum image traces synchrotron emission from
ultra-relativistic electrons accelerated by the expanding supernova
shock wave, and marks the extent of the forward shock \citep{Amy93}.
In young SNRs, strong radio continuum emission can also be found in the
contact region between the forward and reverse shocks \citep{DeLaney04}.
The IR emission associated with young SNRs can originate from several
different mechanisms (see Section~\ref{s:IR-origin} for discussion).

The comparison between the 24 $\mu$m and X--ray images of E0102
reveals surprisingly similar structures despite the very different
wavelengths and emission mechanisms.
The {\it SST} 24 $\mu$m image of E0102 overlaid with the
{\it Chandra} X--ray contours from \cite{Gaetz00} is shown in
Figure~\ref{f:24m_Xray}a.  The position and curvature of the two IR
knots agree remarkably well with those seen in the X--rays.
The ``spoke'' has a very similar size in both the IR and X--ray data
while the bright arc is more confined in the IR.
The region on the northeast side has the lowest 
intensity at 24 $\mu$m, again in agreement with X-ray observations.  
Despite these similarities, some differences are also apparent.  
For example, while the X--ray image has a
high contrast, edge--brightened ring structure, the 24 $\mu$m image
displays a lower contrast between the ring and the central region. 
In addition, the intensity
of the IR emission in the limb does not decrease as sharply towards the
edge of the SNR as the X-ray emission does.

The morphology of the IR emission is less similar to that seen in 
the radio continuum distribution.
Figure ~\ref{f:24m_Xray}b shows a comparison of the 24 $\mu$m emission
and the 6 cm image by \cite{Amy93}.  The IR emission
extends all the way to the edge of the radio continuum distribution. 
Only in the southwestern part is the radio continuum
emission significantly more extended than the IR emission. 
The IR knots do not have corresponding features in the radio continuum.

\begin{figure*}
\epsscale{1.5}
\plotone{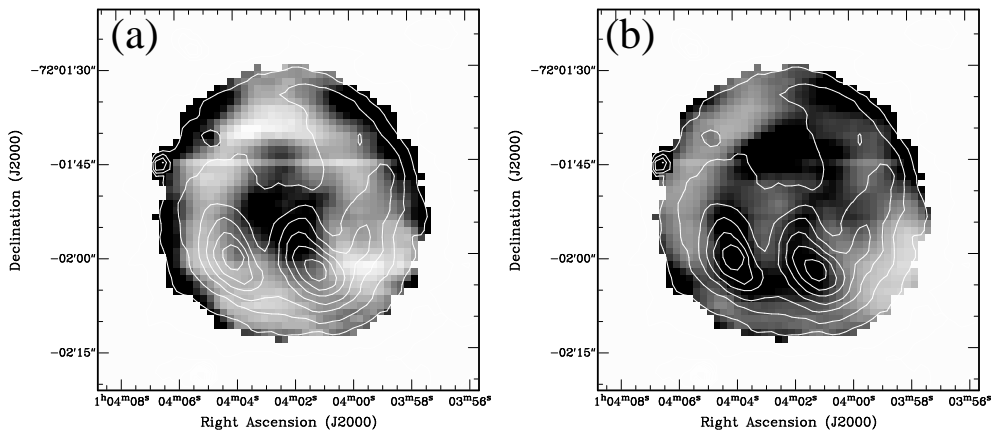}
\caption{\label{f:ratios}   (a) The ratio of
normalized 24 $\mu$m and X--ray images.  The grey-scale intensity
ranges from 0.5 (white) to 2.0 (dark), with a linear transfer
function. Contours, for guidance, are
from the 24 $\mu$m image and range over 20--90\% of the peak value,
with a step of 10\%. (b) The ratio image of
the normalized 24 $\mu$m and 6 cm images. The grey-scale intensity
ranges from 0.5 to 2.0.}
\end{figure*}

For a more quantitative comparison we convolved and 
regridded the X--ray and 6 cm images to match the resolution and 
pixel size of the 24 $\mu$m image.
We then normalized each image to its peak value 
(each normalized image has a peak value equal to 1).
The normalization factors were: 4.7 MJy sr$^{-1}$, $1.1\times10^{-3}$
counts arcsec$^{-2}$ s$^{-1}$, and 177 mJy beam$^{-1}$, 
for the 24 $\mu$m, X--ray, and 6 cm images, respectively. 
Three ratio images were then derived from the normalized images: 
24 $\mu$m/X--ray (Figure~\ref{f:ratios}a), 
24 $\mu$m/6 cm (Figure~\ref{f:ratios}b), and 6 cm/X--ray (not shown).
Figure~\ref{f:ratios}a emphasizes the good spatial correlation between 
the 24 $\mu$m emission and the bright X--ray ring.
The almost featureless white ring has a typical pixel value of $\sim0.7-1$. 
The bright arc is almost invisible in this ratio image, while only the
north part of the ``spoke'' is noticeable, with a slightly higher
ratio of $\sim1.2$. The SNR's edges are mainly dark, with a higher ratio
($>1.5$), emphasizing that the 24 $\mu$m distribution 
decreases less sharply than the X--ray distribution. 
The striking feature in this figure is the central region of the
SNR, particularly  between the two IR knots,
which has the highest IR intensity relative to the X--ray 
intensity, $\sim2$.  
At the same location in the IR/radio ratio image 
(Figure~\ref{f:ratios}b) there is 
a relatively low ratio ($\sim1$), indicating enhanced synchotron emission (at 
6 cm). The ratio image in  
Figure~\ref{f:ratios}b  shows a thinner ring with a larger radius, 
reaching the edge of the forward shock. 
This image has many more significant features, emphasizing 
the generally poor spatial correlation between the IR and radio 
continuum distributions. The two knots are prominent in this image as their IR
intensity is in significant excess relative to the radio continuum intensity.
The region on the northeast side of the SNR with the lowest
intensity at 24 $\mu$m stands out in this ratio image as it 
appears faint in the radio relative to the IR.

\section{Discussion}
\label{s:discussion}

\subsection{Origin of IR emission}
\label{s:IR-origin}

E0102 is the first SNR in the SMC detected in the IR, and it is a
prominent object at 24 $\mu$m.  
Our comparison of the 24 $\mu$m image with radio and X--ray 
observations suggests that the IR emission is 
associated with both forward and reverse shocks.
The noticeable excess of the IR emission relative to the
bright X--ray ring suggests the presence of thermal dust continuum
emission associated with the SNR and originating from the
circumstellar and/or interstellar dust being heated by the X--ray
emitting plasma.
For collisionally heated interstellar dust grains, with a
grain radius of 0.01 $\mu$m and immersed in a hot gas 
with $n\sim1-10$ cm$^{-3}$ and $T\sim10^{7}$ K, 
the expected dust temperature is
$T_{d}\ga130$ K (based on Dwek 1987). 

However, circumstellar or interstellar dust grains embedded in the hot
SN cavity are destroyed  by sputtering and their lifetime is given
by $\tau_{\rm sput}\approx 10^{6} a/n$ yr, where $n$ is the
density of the hot post-shock cavity (in cm$^{-3}$), $a$ is the
radius of dust grains (in $\mu$m), and the assumed temperature is
$\ga10^{6}$ K \citep{Dwek81}.  Assuming a typical size for small
dust grains in the SMC $a=0.005-0.05$ $\mu$m \citep{Stanimirovic00} 
and knowing that the
grains have to survive for the age of E0102 (10$^{3}$ yr),
we estimate $n\la5-10$ cm$^{-3}$ for the 
density of the post-shock cavity. 
This is in agreement with estimated densities from
modeling UV/optical emission lines \citep{Blair00}.

\subsection{Hot dust in 1E 0102.2-7219}

Our measured flux density at 24 $\mu$m and upper limits at 8 and 70 $\mu$m
suggest that the IR emission peaks around 24 $\mu$m and sharply
falls off toward shorter and longer wavelengths. 
This is suggestive of the existence of hot dust associated with the
SNR, with a temperature of $T_{d}\sim120$ K.  
In this scenario, our 1-$\sigma$ upper limits at 8 and 70 $\mu$m are
slightly higher than the expected IR flux densities 
for a quasi-blackbody spectrum at
$T_{d}\sim120$ K ($\propto \nu^{2} B_{\nu}(T_{d})$).  
We have taken into account the contribution from 
the synchrotron emission, extrapolated from
radio measurements \citep{Payne04}, which
is not significant (only $\sim1$ mJy at 24 $\mu$m).
The estimated temperature of hot dust in E0102 is in accord with
expectations for the collisionally heated interstellar and/or
circumstellar dust.

However, the two bright X--ray knots and a significant
fraction of the bright X--ray ring structure correlate 
spatially very well with the IR distribution. This suggests that a
fraction of IR emission is most likely associated with the 
reverse shock as well, 
and may even be largely due to line emission.  
The MIPS band at 24 $\mu$m contains emission lines of [OIV] at 25.88
$\mu$m and [FeII] at 25.98 $\mu$m.  Both lines were observed in 
the young Galactic SNRs RCW103 \citep{Oliva99} and Cas A \citep{Arendt99}.  
The fact that E0102 is detected only at 24 $\mu$m, as well as the fact that
this is the only SNR we detect in these {\it SST} observations 
(for discussion see Bolatto et al., in preparation), also argue in
favor of a significant contribution from emission lines of  
[OIV] and/or [FeII].
As Fe is found to be almost absent, to a level of $\sim10^{-2}-10^{-3}$
relative to O, in the ejecta of E0102 \citep{Hayashi94,Blair00}
the line emission of [OIV] is more likely to affect our 24 $\mu$m measurements.
Future {\it SST} spectroscopic observations (J. Rho, private
communication) will be able to tell
us the exact contribution of [OIV] to the IR flux at 24 $\mu$m.  
For a preliminary estimate 
we use the 24 $\mu$m/X--ray ratio image (Figure~\ref{f:ratios}a)
and assume that all the IR emission in excess to X--ray emission
(pixels with values $>1-1.2$, mainly in the central region and at the
SNR's edges) correspond to the dust continuum emission.
All other IR emission (IR emission that correlates
well with the X--ray emission) is predominantly due to [OIV] emission.
By integrating the 24 $\mu$m image over these two exclusive subsets, 
we estimate that $\la60$\% of the total IR emission from E0102
could be due to the line emission.

If we assume the dust mass absorption coefficient
$\kappa_{\lambda}=2.5(\lambda/{\rm 450}\mu m)^{-2.0}$ 
cm$^2$ g$^{-1}$ \citep{Draine84}
and a distance of 60 kpc \citep{Westerlund91}, we estimate 
the total mass of hot dust associated with the SNR to be 
$M_{d}=8\times10^{-4}$ M$_{\odot}$.
As shown above, up to 60\% of the 24 $\mu$m flux could be due to the line
emission.  This gives a lower limit on the mass of hot dust associated
with the SNR, $M_{d}\ga3\times10^{-4}$ M$_{\odot}$.
Even if {\it all} the hot dust traced at 24 $\mu$m was formed in the
explosion of E0102, the estimated mass of dust is 
significantly lower than what is expected from some recent 
theoretical models. For example, \cite{Todini01}
suggested that a large amount of dust could be formed in explosions of
core-collapse SNRs even in early galaxies with very low metallicity; 
for a progenitor star with a mass of 12--35 M$_{\odot}$
this model predicts $0.08<M_{d}<0.3$ M$_{\odot}$. 
This range is further predicted to increase 
by a factor of 2--3 as the metallicity approaches the solar value.  
With the metallicity of $\sim1/10$ the solar value, 
the amount of dust we estimate for E0102 is at least 
a factor of 100 lower than the lowest predicted value. 
It is interesting, though, that the amount of
hot dust in E0102 is only a few times lower than 
that found in the recent measurement of hot dust in Cas A \citep{Hines04}, 
although the two SNRs occurred in very
different interstellar environments.

\section{Conclusions}
We have obtained IR observations of the young, oxygen-rich SNR 
E0102 as a part of a
large program to image the SMC with the {\it SST} in all available
bands.  The remnant is detected only at 24 $\mu$m, and its IR
distribution has a filled morphology with two prominent
elongated filaments. Our comparison of IR observations with the
existing X--ray and radio continuum images suggests that the IR emission 
is most likely associated with both reverse and forward shocks. 
We find evidence for the existence of up to $8\times10^{-4}$ M$_{\odot}$ of
hot dust at a temperature of $\sim120$ K, associated with the remnant.
The [OIV] emission line may be contributing significantly 
(up to 60\%) to our flux measurement at 24 $\mu$m. 
The hot dust is mainly located in the central region of E0102 which
appears significantly enhanced in IR and radio continuum 
emission relative to the X--ray emission.
The amount of dust associated with E0102, under the 
assumption that {\it all} the dust was formed
during the SN explosion, is at least 100 times lower than what is
predicted by some recent theoretical models for dust production in the
ejecta of core-collapse supernovae.

\begin{acknowledgements}
We thank Carl Heiles for stimulating discussions and 
an anonymous referee for valuable suggestions.
This work is based [in part] on observations made with the 
{\it Spitzer Space Telescope}, which is operated by the Jet Propulsion
Laboratory, California Institute of Technology under NASA contract
1407. Support for this work was provided by NASA through 
Contract Number 1264151 issued by JPL/Caltech.
We also acknowledge support by NSF grants AST-0097417 and AST-0228963.
\end{acknowledgements}

\label{lastpage}

\end{document}